\documentclass[12pt]{iopart}
\usepackage{color}

\newcommand{\lp}{\left(}
\newcommand{\rp}{\right)}

\newcommand{\lsb}{\left[}
\newcommand{\rsb}{\right]}

\newcommand{\labs}{\left|}
\newcommand{\rabs}{\right|}

\newcommand{\abs}[1]{\labs #1 \rabs}

\newcommand{\KAISTNQe}{Department of Nuclear and Quantum Engineering, KAIST, Daejeon 34141, Korea}
\newcommand{\JET}{EUROfusion Consortium, JET, Culham Science Centre, Abingdon, OX14 3DB, UK}
\newcommand{\IPPGreifswald}{Max-Planck-Institut f\"{u}r Plasmaphysik, 17491 Greifswald, Germany}
\newcommand{\CCFE}{Culham Centre for Fusion Energy, Culham Science Centre, Abingdon OX14 3DB, UK}

\usepackage{iopams}
\usepackage{graphicx}
\usepackage{epstopdf}

\newcommand{\refeq}[1]{Eq. (\ref{#1})}
\newcommand{\refsec}[1]{Sec. \ref{#1}}

\newcommand{\reffig}[1]{Fig. \ref{#1}}
\newcommand{\reftable}[1]{Table \ref{#1}}

\begin{document}

\title[]{Bayesian electron density inference from JET lithium beam emission spectra using Gaussian processes}

\author{Sehyun~Kwak$^{1,2\dagger}$, J.~Svensson$^{2,\ddagger}$, M.~Brix$^{3}$, Y.-c.~Ghim$^{1\ast}$, and JET Contributors\footnote{See the Appendix of F. Romanelli et al., Proceedings of the 25th IAEA Fusion Energy Conference 2014, Saint Petersburg, Russia}}

\address{\JET}
\address{$^{1}$ \KAISTNQe}
\address{$^{2}$ \IPPGreifswald}
\address{$^{3}$ \CCFE}
\ead{$^\dagger$slayer313@kaist.ac.kr, $^\ddagger$jakobsemail@gmail.com, $^\ast$ycghim@kaist.ac.kr}
\vspace{10pt}
\begin{indented}
\item[]\today
\end{indented}

\begin{abstract}
A Bayesian model to infer edge electron density profiles is developed for the JET lithium beam emission spectroscopy (Li-BES) system, measuring Li I (2p-2s) line radiation using 26 channels with $\sim1$ cm spatial resolution and $10\sim20$ ms temporal resolution. The density profile is modelled using a Gaussian process prior, and the uncertainty of the density profile is calculated by a Markov Chain Monte Carlo (MCMC) scheme. From the spectra measured by the transmission grating spectrometer, the Li I line intensities are extracted, and modelled as a function of the plasma density by a multi-state model which describes the relevant processes between neutral lithium beam atoms and plasma particles. The spectral model fully takes into account interference filter and instrument effects, that are separately estimated, again using Gaussian processes. The line intensities are inferred based on a spectral model consistent with the measured spectra within their uncertainties, which includes photon statistics and electronic noise. Our newly developed method to infer JET edge electron density profiles has the following advantages in comparison to the conventional method: i) providing full posterior distributions of edge density profiles, including their associated uncertainties, ii) the available radial range for density profiles is increased to the full observation range ($\sim26$ cm), iii) an assumption of monotonic electron density profile is not necessary, iv) the absolute calibration factor of the diagnostic system is automatically estimated overcoming the limitation of the conventional technique and allowing us to infer the electron density profiles for all pulses without preprocessing the data or an additional boundary condition, and v) since the full spectrum is modelled, the procedure of modulating the beam to measure the background signal is only necessary for the case of overlapping of the Li I line with impurity lines.
\end{abstract}

%
\vspace{2pc}

\submitto{\NF}

%
%

\section{Introduction}
\label{sec:intro}
Edge electron density profiles have been recognised as one of the key physical quantities in magnetic confinement devices for controlling and understanding edge plasma phenomena, such as edge localised modes (ELMs) \cite{96PPCF_Zohm}, L-H transitions \cite{84PRL_Wagner} and turbulence transport \cite{97IEEE_Carreras}. Lithium beam emission spectroscopy (Li-BES) systems, capable of providing the profiles of edge electron density, have thus been widely used at various devices (TEXTOR \cite{92PPCF_Schweinzer, 93RSI_Wolfrum}, ASDEX Upgrade \cite{97FED_McCormick, 08PPCF_Fischer}, W7-AS \cite{97FED_McCormick}, and JET \cite{93PPCF_Pietrzyk, 10RSI_Brix, 12RSI_Brix}). Li-BES system is a type of beam diagnostics that injects neutral lithium atoms into the plasma and measures Li I (2p-2s) line radiation caused by spontaneous emission processes from the first excited state (1s2 2p1) to the ground state (1s2 2s1) of the neutral lithium beam atoms. The Li I line intensity can be expressed as a function of the plasma density by a multi-state model \cite{91APB_Schorn} which describes the relevant processes between lithium atoms and plasma particles. The profiles of edge electron density can be inferred from the measured profiles of the Li I line intensity.

The integral expression of the multi-state model which calculates a profile of electron density \cite{92PPCF_Schweinzer} from the measured Li-BES data has been used conventionally at many devices \cite{93RSI_Wolfrum, 97FED_McCormick, 93PPCF_Pietrzyk, 10RSI_Brix}. This method, however, has a limitation that profiles of absolute electron density (based on the absolute calibration factor) can be obtained only if either a \textit{singular point} is found or an additional boundary condition is provided in the data. Consequently, this method involves some weaknesses: i) preprocessing of the data is usually required to find the singular point, ii) the singular point cannot be found accurately, iii) a small change of the location of the singular point can cause a large difference of the density profile and iv) an additional boundary condition, which is required if the singular point does not exist, cannot be properly fixed because of the  difficulty of obtaining all the populations of the different states of the neutral Li beam atoms. Another method utilising Bayesian probability theory to analyse the Li-BES data was reported at ASDEX Upgrade \cite{08PPCF_Fischer}, using non-spectral APD (Avalanche Photo Diode) detectors and made impressive progress. Our method fits the full Li beam emission spectrum and uses Gaussian processes to model and regularise the electron density profiles, rather than using the non-spectral data and the combination of splines with a regularising \textit{weak} monotonicity constraint used in \cite{08PPCF_Fischer}. Our proposed method requires neither preprocessing of the data, inner boundary information nor a profile monotonicity regulariser.

The method comprises two parts. The first part is obtaining the profile of the Li I line intensity. The JET Li beam emission spectrum is here modelled as a single Li I emission line and a background signal, convolved with an instrument function and filtered through an interference filter. The interference filter and instrument function need to be separately estimated and the noise on the spectra is modelled by an electronic offset as well as photon statistics and electronic noise. We infer interference filter and instrument functions based on separate measurements (which are required only once in a while as they do not vary much shot-to-shot base) using Gaussian processes. We use Gaussian processes because we do not know the parametric form, i.e., analytical expression of these functions. Having the interference filter and instrument functions, we then infer intensities of Li I line radiation, background and the electronic offset simultaneously. This provides the advantage of removing the necessity of beam modulations to obtain separate background measurements within a plasma shot. Furthermore, as the fitted background intensity is likely to be dominated by Bremsstrahlung radiation, our method opens a possibility to obtain the effective charge $Z_\mathrm{eff}$. The second part of our method infers the profile of edge electron density based on the intensity profile of Li I line radiation using the multi-state model. During this second part, the absolute calibration factor of the system is inferred directly from the measurements, removing the need for the singular-point method mentioned above. All modelling and analyses are performed using a Bayesian scheme within the Minerva framework \cite{07ISP_Svensson}.

\refsec{sec:model} describes the models we use: the multi-state model describing how to obtain electron density information from the Li I line radiation intensity and the spectral model of the raw data, forming together the forward model of the JET Li-BES system. \refsec{sec:inference} explains how the interference filter and instrument functions are inferred and the procedure for obtaining the intensity of the Li I line radiation and electron density profile. Conclusions are presented in \refsec{sec:conclusion}.

\section{Models}
\label{sec:model}
\subsection{Multi-state model}
\label{sec:model-physics}
Li-BES system measures the intensities of the Li I (2p-2s) line radiation from the neutral lithium beam penetrating into the plasma. The Li I line radiation is produced by spontaneous emission processes from the first excited state (1s2 2p1) to the ground state (1s2 2s1) of the neutral lithium beam atoms. The Li I line intensity is a function of a population of the first excited state which can be expressed in terms of a plasma density via a multi-state model.

The change of relative populations in time using the multi-state (collisional-radiative) model \cite{92PPCF_Schweinzer} is
\begin{equation}
\frac{dN_i\lp t\rp}{dt}=\sum\limits_{j=1}^{M_{\mathrm{Li}}}\lsb\sum\limits_{s}n_sa^s_{ij}\lp v^s_r\rp+b_{ij}\rsb N_j,
\label{eq:multi-state-time}
\end{equation}
which describes population and de-population of states of the neutral lithium atoms caused by processes between lithium beam atoms and plasma particles in addition to spontaneous emissions. $N_i$ is a \textit{relative} population of the $i^\mathrm{th}$ state with respect to the total number of the neutral lithium beam atoms at the position where the lithium beam enters the vacuum vessel. For instance, $N_1=0.7$ and $N_2=0.1$ mean that $70$ \% and $10$ \% of the initial neutral lithium beam atoms are in the ground and first excited states, respectively. $M_{\mathrm{Li}}$ is the number of states of the neutral lithium atoms, and we consider nine different states in this paper; thus, $M_{\mathrm{Li}}=9$. $n_s$ is a plasma density of species $s$ where $s=e$ and $s=p$ denote electron and proton, respectively. $a^s_{ij}\lp i\neq j\rp>0$ is a net population rate coefficient by the plasma species $s$ from the $j^\mathrm{th}$ state to the $i^\mathrm{th}$ state increasing the relative population of the $i^\mathrm{th}$ state, while $a^s_{ii}<0$ is a net de-population rate coefficients including excitation, de-excitation and ionisation effects leaving the $i^\mathrm{th}$ state. All population and de-population rate coefficients caused by plasma species $s$ depend on the relative speed between the neutral lithium beam atoms and plasma species $s$ which is denoted as $v^s_r$. $b_{ij}$ is the spontaneous emission rate coefficient or Einstein coefficient.

It becomes easier to solve \refeq{eq:multi-state-time} if it is expressed in terms of the beam coordinate $z$: $d/dt=d/dz\cdot dz/dt$. Realising that $dz/dt$ is the velocity of the neutral lithium beam atoms $v_{\mathrm{Li}}$, we obtain
\begin{eqnarray}
\frac{dN_i\lp z\rp}{dz}&=&\frac{1}{v_{\mathrm{Li}}}\sum\limits_{j=1}^{M_{\mathrm{Li}}}\lsb\sum\limits_{s}n_s\lp z\rp a^s_{ij}\lp v^s_r\lp z\rp\rp+b_{ij}\rsb N_j\lp z\rp.
\label{eq:multi-state-beam}
\end{eqnarray}
Here, we assume that $v_{\mathrm{Li}}$ is constant over the penetration range of the beam into plasmas.

\begin{figure}
\includegraphics[width=\linewidth]{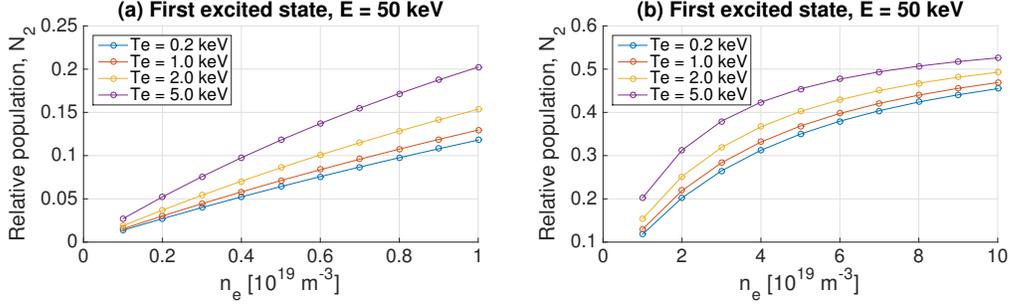}
\caption{Steady-state relative populations of the first excited state ($N_2$) of the neutral lithium beam atoms as a function of the electron density and temperature with a beam energy of $50$ keV in the range of (a) $0.1\times 10^{19}<n_e<1.0\times 10^{19}$ and (b) $1.0\times 10^{19}<n_e<10.0\times 10^{19}$.}
\label{fig:atomicdata}
\end{figure}

The relative speed $v^s_r\lp z\rp$ is not directly measured but can be approximated using other quantities. The relative speed between the neutral lithium beam atoms and electrons $v^e_r\lp z\rp$ is dominated by the electron temperature $T_e$ since the typical (thermal) speed of electrons is much faster than that of the neutral lithium beam atoms. The relative speed between the neutral lithium beam atoms and protons $v^p_r\lp z\rp$ can be approximated to the lithium beam velocity in case of JET Li-BES since the lithium beam energy is $\sim55$ keV which is much higher than the ion temperature. Other species are not considered in this work. Thus, the multi-state model becomes
\begin{eqnarray}
\frac{dN_i\lp z\rp}{dz}=\frac{1}{v_{\mathrm{Li}}}\sum\limits_{j=1}^{M_{\mathrm{Li}}}\lsb n_e\lp z\rp a^e_{ij}\lp T_e\lp z\rp\rp+n_p\lp z\rp a^p_{ij}\lp v_{\mathrm{Li}}\rp+b_{ij} \rsb N_j\lp z\rp,
\label{eq:multi-state}\\
N_i\lp z=0\rp=\delta_{1i},
\label{eq:initial}
\end{eqnarray}
with the initial condition \refeq{eq:initial} where we assume that all the lithium beam atoms are neutral and in the ground state ($i=1$) at the initial position where the beam enters the tokamak vacuum vessel corresponding to $z=0$, i.e., $N_1\lp z=0\rp = 1$. The rate coefficients have been obtained from the Atomic Data Analysis Structure (ADAS) \cite{ADAS} and the reference \cite{99ADNDT_Schweinzer}. \reffig{fig:atomicdata} shows an example of steady-state relative populations for the first excited state $N_2$ as a function of electron density and temperature with a beam energy of $50$ keV.

Note that this multi-state model does not consider the population of ionised lithium atoms, which leave the beam due to a strong magnetic field of JET. Therefore, electron loss processes such as ionisation and charge-exchange simply attenuate the total population of the neutral lithium beam atoms, i.e., $\sum\limits_{j=1}^{M_{\mathrm{Li}}} N_j\lp z>0\rp <1$.

\subsection{Spectral model}
\label{sec:model-spectral}
The JET Li-BES system measures spectra, including the Doppler shifted Li I line radiation from the 26 different spatial positions, covering a few nanometres in wavelength using the transmission grating spectrometer (dual entrance slit with interference filter for preselection of passband, details in \cite{12RSI_Brix}). A charge coupled device (CCD) camera detects the photons for integration time of approximately 10 ms. More detailed description of the JET Li-BES system can be found elsewhere \cite{10RSI_Brix, 12RSI_Brix}. 

A spectrum from each spatial position contains four types of signals (in addition to noise): i) Li I line, ii) a background dominated by Bremsstrahlung radiation, iii) an electronic offset and iv) impurity lines. Doppler broadening of the Li I line radiation is negligible since the lithium beam is a mono-energetic beam ($\sim0.02$ nm broadening occurs for the beam temperature of $\sim10$ eV, and the dispersion of the CCD pixel is $\sim0.04$nm/pixel), therefore we treat the Li I line as a delta function in the spectrum. A measured spectrum $S\lp\lambda\rp$ from each spatial position can be expressed as
\begin{equation}
S\lp\lambda\rp=F\lp\lambda\rp\lsb C\lp\lambda\rp A+B\rsb+Z,
\label{eq:spectrum}
\end{equation}
where $A$ is the intensity of Li I line radiation, $B$ the background level and $Z$ the electronic offset, which are all inferred together with their uncertainties through Bayesian inference. The instrument function $C\lp\lambda\rp$ and interference filter function $F\lp\lambda\rp$ are inferred through a Bayesian scheme using Gaussian processes from separate measurements \cite{15RSI_Sehyun}. Here, $\lambda$ is the wavelength corresponding to a CCD pixel index \cite{10RSI_Brix}.

Gaussian processes are probabilistic functions defined by a multivariate Gaussian distribution whose mean and covariance function specifies the mean and the covariance between any two points in the domain \cite{06MIT_Carl}. This constrains the variability of the function without any analytic specification, i.e., in a non-parametric way. Gaussian processes were introduced in the fusion community in \cite{11EFDA_Svensson} and are implemented as a standard representation of profile quantities in the Minerva framework \cite{07ISP_Svensson}. It has been used for current tomography \cite{11EFDA_Svensson,13NF_Romero}, soft x-ray tomography \cite{13RSI_Li}, and representing profile quantities \cite{11EFDA_Svensson,11EPS_Schmuck,15NF_Chilenski}. The covariance function of a Gaussian process is defined as a parametrised function whose parameters, so called \textit{hyperparameters}, determine aspects of the function such as overall scale and length scale. The hyperparameters are selected based on the measurements by maximising the \textit{evidence} through Bayesian model selection. A detailed description of the Bayesian inference and modelling of the JET Li-BES data with Gaussian process can be found elsewhere \cite{15RSI_Sehyun}.

\subsection{Forward model}
\label{sec:forward}
\begin{figure}
\includegraphics[width=\linewidth]{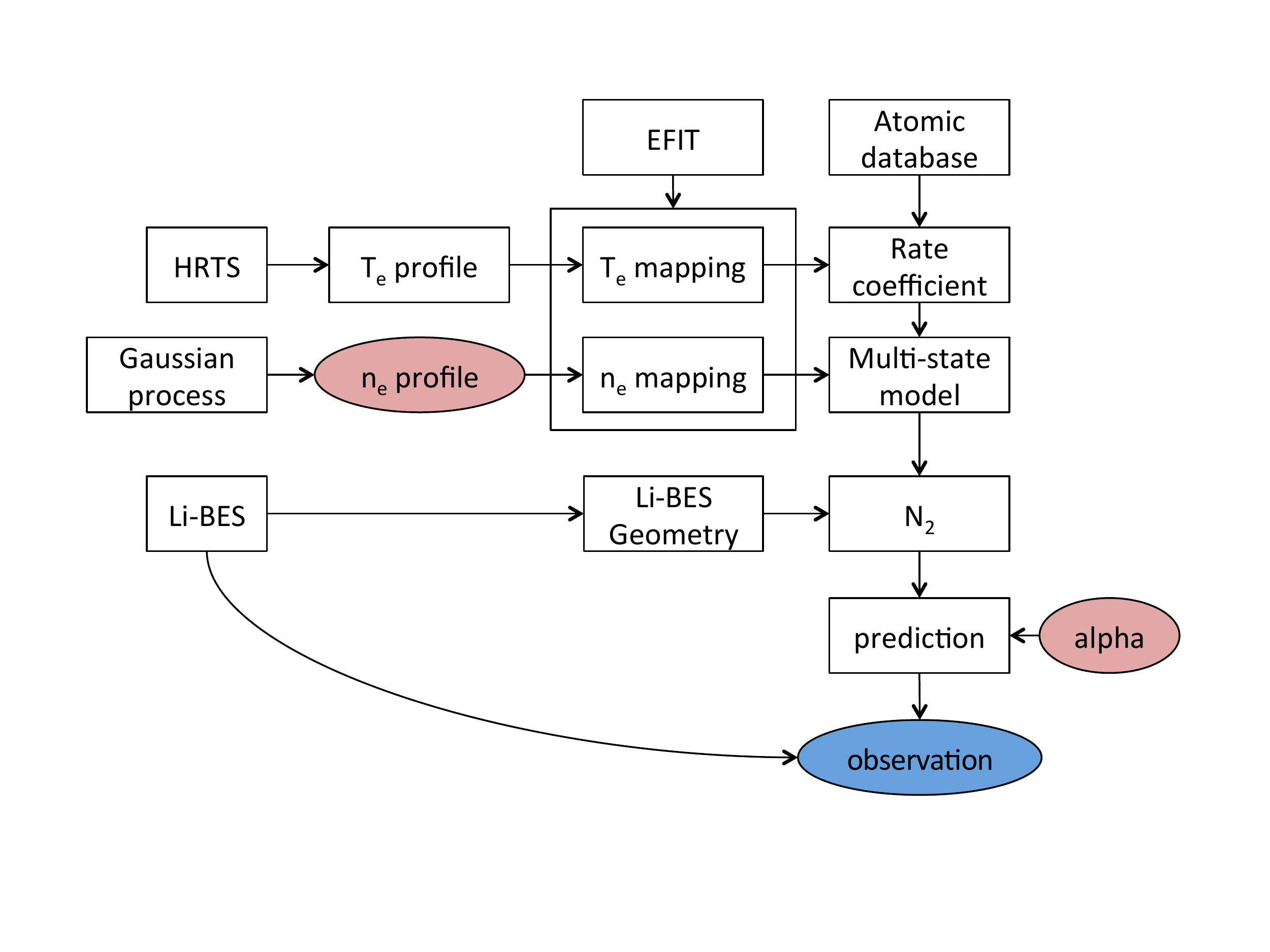}
\caption{A simplified graphical representation of the JET Li-BES forward model as implemented in the Minerva Bayesian modelling framework \cite{07ISP_Svensson}. The free parameters are shown with red circles and observations as a blue circle. The rectangular boxes represent operations or constants. The electron density $n_e$ and temperature $T_e$ are mapped onto the EFIT estimated flux surfaces. The relative populations of the neutral lithium beam atoms are calculated from the multi-state model, and profiles of the Li I line radiation intensities are predicted given edge $n_e$ profiles and an absolute calibration factor, alpha ($\alpha$). All the possible edge $n_e$ profiles whose predicted Li I line intensity profile agree with the observation (blue circle) within their uncertainties are found through a MCMC scheme.}
\label{fig:model}
\end{figure}

Our goal is to find all possible profiles of the edge electron density $n_e$ consistent with the spectral observations. For this, we consider the forward model as shown in \reffig{fig:model}. The edge electron density profile $n_e$ is modelled as a set of values at given positions, with a prior given by a Gaussian process with given overall scale and scale length hyperparameters, discussed in more detail in \refsec{sec:inference-density}. Edge $n_e$ profiles are mapped onto flux surface coordinates $\psi$ calculated by the EFIT equilibrium code. Electron temperature $T_e$, required for the rate coefficients $a_{ij}^s$, is measured by the High Resolution Thomson Scattering (HRTS) system \cite{04RSI_Pasqualotto} and mapped onto the same flux surface coordinates. This will allow us to calculate a relative population of the first excited state of the neutral lithium beam atoms, i.e., $N_2$, based on the multi-state model \refeq{eq:multi-state} with a quasi-neutrality condition, i.e., $n_e=n_p$. Here, we assume that impurity densities are low enough to be ignored\footnote{If impurities are non-negligible, then our measured spectra may show strong impurity line radiation in which case our assumption is not valid.}. 

Once we have $N_2$, we can predict the Li I line radiation intensity, $A$ in \refeq{eq:spectrum}, where the detailed procedure is provided in \refsec{sec:model-forward}. This model provides a prediction of the measured Li I line radiation $A_*$, given the free parameters of an electron density $n_e$ and an absolute calibration factor $\alpha$, by
\begin{equation}
p\lp A_* | n_e, \alpha\rp = \frac{1}{\sqrt{2\pi\sigma}}\exp\lsb{-\frac{\lp A_*-A\lp n_e, \alpha\rp\rp^2}{2\sigma^2}}\rsb,
\label{eq:simple_likelihood}
\end{equation}
where $A\lp n_e, \alpha\rp$ is a model prediction with specific values of the free parameters, $n_e$ and $\alpha$. $\sigma$ is the uncertainty associated with the observation $A_*$. This is our basic form of the forward model in this paper and is the \textit{likelihood} in Bayes formula (\refeq{eq:Bayes}). We assume that deviations of the observation from predictions have a Gaussian distribution. We discuss how we estimate $\sigma$ and rationale to form Gaussian distributed deviations in \refsec{sec:model-uncertainties}.

\subsubsection{Detected number of photons}
\label{sec:model-forward}
The Li I line intensity, $A$ in \refeq{eq:spectrum}, is proportional to the relative population of the first excited state $N_2$, i.e., $A\propto b_{12}N_2$ where $b_{12}$ is the spontaneous emission rate coefficient from the first state to the ground state. Change of the relative population of the first excited state due to the spontaneous emission as the beam travels a distance of $\Delta z$ denoted as $\abs{\Delta N_2}$ is
\begin{equation}
\abs{\Delta N_2\lp z\rp}=\frac{\abs{b_{12}}}{v_{\mathrm{Li}}}\Delta z\ N_2\lp z\rp,
\label{eq:fraction}
\end{equation}
where $\Delta z$ can be considered as the observation length. Since one spontaneous emission produces one photon, the total number of emitted photons $N_{\mathrm{ph}}^\mathrm{em}$ corresponding to Li I line radiation over the integration time $\Delta t$ with the lithium beam current $I_{\mathrm{Li}}$ is
\begin{equation}
N_{\mathrm{ph}}^\mathrm{em}\lp z\rp=I_{\mathrm{Li}}\Delta t\abs{\Delta N_2\lp z\rp}=I_{\mathrm{Li}}\Delta t \frac{\abs{b_{12}}}{v_{\mathrm{Li}}}\Delta z\ N_2\lp z\rp.
\label{eq:photons}
\end{equation}
The emitted photons falling into the solid angle of the collection optics pass through various mirrors, lens and grism before being detected by the CCD camera. We denote all these effects of optics including the solid angle as an effective transmittance of the system, $T$. Then, the number of photons detected by (or arrived to) the CCD camera $N_\mathrm{ph}^\mathrm{det}\lp z\rp$ is 
\begin{equation}
N_\mathrm{ph}^\mathrm{det}\lp z\rp=T N_{\mathrm{ph}}^\mathrm{em}\lp z\rp=T I_{\mathrm{Li}}\Delta t \frac{\abs{b_{12}}}{v_{\mathrm{Li}}}\Delta z\ N_2\lp z\rp.
\label{eq:arrived_photons}
\end{equation}
Also, we define $Q$ as the count per photon of the CCD camera. $Q$ describes the number of counts produced by the CCD camera when one photon arrives at the CCD detector. Then, the CCD output count due to the Li I line radiation $N_\mathrm{CCD}^\mathrm{Li}$ which we measure is
\begin{eqnarray}
N_\mathrm{CCD}^\mathrm{Li}\lp z\rp&=&Q N_\mathrm{ph}^\mathrm{det}\lp z\rp=Q T I_{\mathrm{Li}}\Delta t \frac{\abs{b_{12}}}{v_{\mathrm{Li}}}\Delta z N_2\lp z\rp
\nonumber\\
&=&A\lp z\rp \int F\lp\lambda\rp C\lp\lambda\rp  d\lambda,
\label{eq:counts}
\end{eqnarray}
and this is, by definition, equal to the Li I line intensity $A$ multiplied by the spectrally integrated signal of the instrument function $C\lp\lambda\rp$ and the interference filter function $F\lp\lambda\rp$ in \refeq{eq:spectrum}. 

We finally obtain
\begin{equation}
A\lp z\rp = \underbrace{ \frac{Q T I_{\mathrm{Li}}\Delta t \frac{\abs{b_{12}}}{v_{\mathrm{Li}}}\Delta z}{\int F\lp\lambda\rp C\lp\lambda\rp  d\lambda}}_{\equiv\alpha} N_2\lp z\rp=\alpha N_2\lp z\rp,
\label{eq:for_alpha}
\end{equation}
where $\alpha$ is the absolute calibration factor which is taken as a free parameter in our forward model in addition to the $n_e$ profile as shown in \reffig{fig:model}. Note that we have included the magnitude of the relative calibration factors in the instrument function $C\lp\lambda\rp$.

\subsubsection{Uncertainties}
\label{sec:model-uncertainties}
The main measurement error is due to the Poisson distributed photon statistics. On top of that, there is an additional electronic noise which is measured before a pulse starts and is here taken as a Gaussian distribution. 

To be able to determine a level of photon noise, it is necessary to find the value of $Q$ in \refeq{eq:counts} so that the measured $N_\mathrm{CCD}^\mathrm{Li}$ can be converted to the detected number of photons $N_\mathrm{ph}^\mathrm{det}$ which is the quantity following a Poisson distribution. With an aim of determining the value of $Q$, we shine a uniform intensity light-emitting diode (LED) to the CCD camera while varying the intensity of the LED with all other conditions fixed as if it were actual measurements of the Li-BES during plasma discharges. The arithmetic mean of CCD output counts $\bar N_\mathrm{CCD}$ and its associated variance $\sigma_\mathrm{CCD}^2$ are
\begin{eqnarray}
\bar N_{\mathrm{CCD}}&=&Q \bar N_{\mathrm{ph}}+\bar N_\mathrm{CCD}^\mathrm{DC}+\bar Z_\mathrm{CCD},
\label{eq:mean-count-photon} \\
\sigma_\mathrm{CCD}^2&=&Q^2\sigma_\mathrm{ph}^2+\sigma_\mathrm{e}^2
\label{eq:var-count-photon}
\end{eqnarray}
where $\bar N_\mathrm{ph}$ is the mean of the number of photons detected by (arrived to) the CCD camera and $\bar N_\mathrm{CCD}^\mathrm{DC}$ the mean CCD output counts due to the dark current of the CCD. Here, $\bar Z_\mathrm{CCD}$ is the mean CCD offset. $\sigma_\mathrm{ph}^2$ and $\sigma_\mathrm{e}^2$ are the variances due to photon statistics and electronic noises, respectively. Note that we treat fluctuations in the dark current as a part of the electronic noise because they exist in the absence of detected photons. 

\begin{figure}
\includegraphics[width=\linewidth]{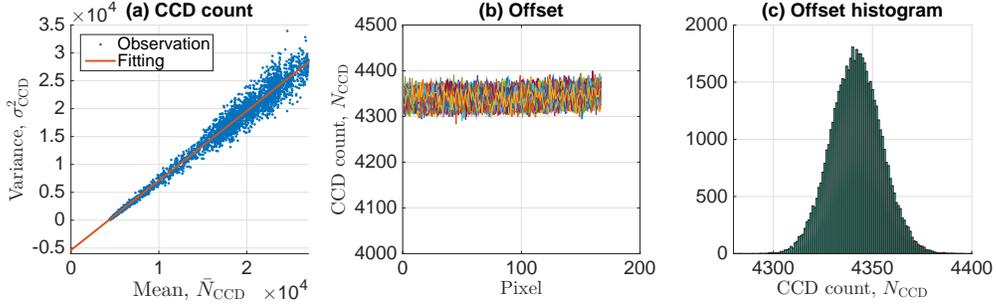}
\caption{ (a) The variance $\sigma^2_{\mathrm{CCD}}$ vs. the mean counts $\bar N_\mathrm{CCD}$ of the CCD output with varying LED intensities. The slope of the fitted linear line is the value of $Q$ which is $1.247\pm0.005$. (b) Measurements of the CCD output counts as a function of the CCD pixel with all the electronics switched on and no input photons to the CCD, and (c) the histogram of the CCD output counts from (b). The histogram shows that i) the variance is 160, i.e., $\sigma_\mathrm{e}^2\approx160$, with the mean value of $4342$ and ii) it has a Gaussian shape.}
\label{fig:count-per-photon}
\end{figure}

With $\bar N_\mathrm{ph}=\lp \bar N_\mathrm{CCD}-\bar N_\mathrm{CCD}^\mathrm{DC}-\bar Z_\mathrm{CCD}\rp/Q$ from \refeq{eq:mean-count-photon} and $\bar N_\mathrm{ph}=\sigma_\mathrm{ph}^2$ owing to a Poisson distribution, recasting \refeq{eq:var-count-photon}, we get
\begin{equation}
\sigma^2_{\mathrm{CCD}}=Q \bar N_{\mathrm{CCD}}-(Q \bar N_\mathrm{CCD}^\mathrm{DC}+Q \bar Z_\mathrm{CCD}-\sigma^2_{\mathrm{e}}).
\label{eq:mean-var-count}
\end{equation} 
Notice that $\bar N_\mathrm{CCD}$ and $\sigma^2_{\mathrm{CCD}}$ can be directly measured with the LED on, and by varying the intensity of the LED we can determine the value of $Q$. \reffig{fig:count-per-photon}(a) shows a graph of the measured $\sigma^2_{\mathrm{CCD}}$ vs. $\bar N_\mathrm{CCD}$, using a total of 4,175 (167 pixels from 25 channels) independent data points, the variances and the means which are estimated using 332 independent time points. The slope is the value of $Q$ we seek, and we find that $Q=1.247\pm0.005$.

To find the electronic noise level $\sigma^2_{\mathrm{e}}$, we switch on all the electronics and measure fluctuations in $N_\mathrm{CCD}$ without any photons to the CCD, i.e., $N_\mathrm{ph}=0$. Here, $N_\mathrm{CCD}$ and $N_\mathrm{ph}$ are individual measurements rather than their means. \reffig{fig:count-per-photon}(b) shows such measurements for all 26 spatial channels (different colours). \reffig{fig:count-per-photon}(c) is the histogram of the $N_\mathrm{CCD}$. The variance is estimated to be $160$ with a mean of $4342$. Therefore, $\sigma_\mathrm{e}^2\approx160$. As can be seen from the histogram, the dark current fluctuations are approximately Gaussian shaped. Furthermore, as we find the mean value of the offset, i.e., $4342$, appears constantly for all channels, we always subtract this offset value from the measured signal before performing any analyses on the data. Any residual offset is captured by $Z$ in \refeq{eq:spectrum}.

When the number of counts is large a Poisson distribution can be approximated with a Gaussian distribution. Since the detected number of photons $N_\mathrm{ph}^\mathrm{det}$ is larger than 100, we take the photon statistics to follow a Gaussian distribution as well. Therefore, the variance $\sigma^2$ in \refeq{eq:simple_likelihood} is 
\begin{equation}
\sigma^2 = \sigma_\mathrm{ph}^2+\sigma_\mathrm{e}^2.
\label{eq:total_variance}
\end{equation}

\section{Bayesian inference}
\label{sec:inference}
For our case, we have a spectrum $S\lp\lambda\rp$ described by three free parameters: the Li I line radiation intensity $A$, the background $B$ dominated by Bremsstrahlung radiation, and the electronic offset $Z$. The instrument function $C\lp\lambda\rp$ and the interference filter $F\lp\lambda\rp$ in \refeq{eq:spectrum} are inferred separately using Gaussian processes. 

In the Bayesian scheme, we calculate the probability distribution of a free parameter $\mathcal{W}$ given observation $\mathcal{D}$ known as the \textit{posterior} $p\lp\mathcal{W}|\mathcal{D}\rp$. The posterior is given by Bayes formula
\begin{eqnarray}
p\lp\mathcal{W}|\mathcal{D}\rp&=&\frac{p\lp\mathcal{D}|\mathcal{W}\rp p\lp\mathcal{W}\rp}{p\lp\mathcal{D}\rp},
\label{eq:Bayes}
\end{eqnarray}
where $p\lp\mathcal{D}|\mathcal{W}\rp$, $p\lp\mathcal{W}\rp$ and $p\lp\mathcal{D}\rp$ are the \textit{likelihood}, \textit{prior} and \textit{evidence}, respectively. The likelihood is a model for observations given free parameters as described in  \refeq{eq:simple_likelihood}. The prior quantifies our assumptions on the free parameters before we have observations. The evidence is typically used for a model selection and is irrelevant if one is only interested in estimating the free parameters. A detailed description of Bayesian inference can be found elsewhere \cite{96Oxford_Sivia}.

To minimise possible confusion, we define our notations used in this section in \reftable{table:notation}. As the JET Li-BES system obtains spectra from 26 different spatial positions, the channel index corresponds to the spatial position and the pixel index to the wavelength. The predicted signal at the $i^\mathrm{th}$ channel and $j^\mathrm{th}$ pixel is denoted as $S^i_j$, and $D^i_j$ represents the observed signal.

\begin{table}\centering
\begin{tabular}{c c p{6cm}} \hline\hline
Expression style & Example & Meaning \\ \hline
Boldface & $\mathbf{y}$, $\mathbf{S}$ & Column vector \\
Boldface with a check accent & $\check\mathbf{K}$, $\check\mathbf{\Sigma}$ & Matrix \\
Plain & $S$, $x$ & Scalar \\
Superscript index & $\mathbf{y}^i$, $x^i$ & Quantity of the $i^\mathrm{th}$ channel out of the total 26 spatial positions (channels) of the JET Li-BES system\\
Subscript index & $\mathbf{y}_i$, $x_i$ & Quantity at the $i^\mathrm{th}$ wavelength in the CCD camera \\ \hline
\end{tabular}
\caption{Notations used in \refsec{sec:inference}.}
\label{table:notation}
\end{table}

Using these notations, we will find the most probable prediction of the line intensity, background and offset at $i^\mathrm{th}$ channel by calculating the posterior $p\lp A^i,B^i,Z^i | \mathbf{D}^i \rp$ where the predicted signal at the $i^\mathrm{th}$ channel and $j^\mathrm{th}$ pixel is
\begin{equation}
S^i_j=F^i_j\lp C^i_j A^i+B^i\rp+Z^i.
\label{eq:spectral-model-vector}
\end{equation}

In the following subsections, we describe how to infer two unknown functions, the interference filter and instrument functions ($\mathbf{F}^i,\mathbf{C}^i$), and the free parameters ($A^i,B^i,Z^i$).

\subsection{Interference filter and instrument functions}
\label{sec:inference-effect}
\begin{figure}
\includegraphics[width=\linewidth]{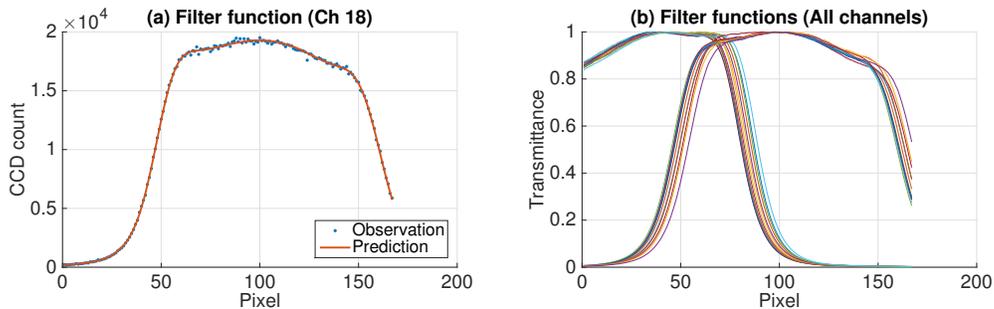}
\caption{ (a) The observation (dots) and the MAP estimate of the filter function (red line) for channel 18 using Bayes formula with the Gaussian process prior, showing a good agreement between the two. (b) Normalised filter functions (MAP) for all channels shown in different colours.}
\label{fig:filter_effect}
\end{figure}

To infer the $i^\mathrm{th}$ channel interference filter function $\mathbf{F}^i$, we illuminate uniform LED light to the fibres. Since there is no Li I line radiation ($A^i$) with a negligible electronic offset ($Z^i$) as shown in \reffig{fig:filter_effect}, the predicted signal is
\begin{equation}
S^i_j=F^i_j\lp C^i_j A^i+B^i\rp+Z^i=F^i_j B^i, 
\label{eq:spectral-model-vector-interference}
\end{equation}
where $B^i$ is uniform LED light intensity. According to Bayes formula, the posterior is 
\begin{equation}
p\lp \mathbf{F}^i | \mathbf{D}^i \rp \propto p\lp \mathbf{D}^i | \mathbf{F}^i \rp p\lp \mathbf{F}^i \rp, 
\label{eq:bayes_spec}
\end{equation}
where the likelihood is
\begin{equation}
p\lp\mathbf{D}^i | \mathbf{F}^i\rp = \frac1{\sqrt{\lp2\pi\rp^{N_{\mathrm{pixel}}}\abs{\check\mathbf\Sigma}}} \exp{\lsb-\frac12\lp\mathbf{D}^i-\mathbf{S}^i\rp^T\check\mathbf\Sigma^{-1}\lp\mathbf{D}^i-\mathbf{S}^i\rp\rsb}.
\label{eq:likelihood}
\end{equation}
Here, $\mathbf{S}^i=\mathbf{F}^i B^i$ as in \refeq{eq:spectral-model-vector-interference}, and $N_{\mathrm{pixel}}$ is the total number of CCD pixels for the $i^\mathrm{th}$ channel. $\check\mathbf\Sigma$ is an $N_{\mathrm{pixel}}\times N_{\mathrm{pixel}}$ square diagonal matrix containing variances of the measured signal at each pixel of the CCD camera as in 
\begin{eqnarray}
\check\mathbf\Sigma&=&
\lsb \begin{array}{cccccc}
\sigma^2_{1} & & & & &\\
& \sigma^2_{2} & & & &\\
& & \dots & & &\\
& & & \sigma^2_{j} & & \\
& & & & \dots & \\
& & & & & \sigma^2_{N_{\mathrm{pixel}}} \\
\end{array}\rsb,
\label{eq:uncertainties}
\end{eqnarray}
where $\sigma^2_j=\sigma^2_{\mathrm{ph},j}+\sigma^2_{\mathrm{e},j}$ at the $j^\mathrm{th}$ pixel as \refeq{eq:total_variance} is used in \refeq{eq:simple_likelihood}. $\sigma^2_{\mathrm{ph},j}$ and $\sigma^2_{\mathrm{e},j}$ can be estimated as described in \refsec{sec:model-uncertainties}. Note that $\check\mathbf\Sigma$ is different for different channels.

The prior $p\lp \mathbf{F}^i \rp$ in \refeq{eq:bayes_spec} needs to be specified. Since we do not know the parametric form, i.e., analytical form, describing the interference filter of the $i^\mathrm{th}$ channel, $\mathbf{F}^i$, as a function of wavelength (pixel index),  we use a Gaussian process prior for $\mathbf{F}^i$:
\begin{equation}
p\lp\mathbf{F}^i\rp = \frac1{\sqrt{\lp2\pi\rp^{N_{\mathrm{pixel}}}\abs{\check\mathbf{K}}}}\exp{\lsb-\frac12\lp\mathbf{F}^i-\mathbf{0}\rp^T\check\mathbf{K}^{-1}\lp\mathbf{F}^i-\mathbf{0}\rp\rsb}.
\label{eq:prior-effect}
\end{equation} 
Here, $\mathbf{0}$ is a column vector whose entries are all zeros. The $N_{\mathrm{pixel}}\times N_{\mathrm{pixel}}$ matrix $\check\mathbf{K}$, which varies channel by channel, is defined as a squared exponential covariance function with the value at the $j^\mathrm{th}$ row and $k^\mathrm{th}$ column of
\begin{equation}
K_{jk} = \sigma_f^2 \exp{\lsb-\frac1{2\ell^2}\abs{x_j - x_k}^2 \rsb} + \sigma_n^2 \delta_{jk}.
\label{eq:cov-effect}
\end{equation}
$\delta_{jk}$ is the Kronecker delta. $\mathbf{x}$ is a vector of the CCD pixel index, thus $\abs{x_j - x_k}$ is the difference in pixel index between the $j^\mathrm{th}$ and $k^\mathrm{th}$ pixels. $\sigma_f^2$ is the signal variance and $\ell$ the scale length. $\sigma_n^2$ is a small number for the numerical stability of the model. The hyperparameters $\sigma_f^2$ and $\ell$  govern the characteristic of the Gaussian process \refeq{eq:prior-effect}, and we find their values by maximising the evidence $p\lp \mathbf{D}^i \rp$. More detailed description can be found elsewhere \cite{15RSI_Sehyun}.

\reffig{fig:filter_effect}(a) shows the comparison between the observation $\mathbf{D}^i$ and the maximum a posteriori (MAP) estimate of $\mathbf{F}^i$ for channel 18. \reffig{fig:filter_effect}(b) shows the MAP estimates of the filter functions for all channels of the JET Li-BES system. Note that we normalise all the filter functions to have the maximum value of one as what we need is the \textit{shape} of the filter functions in the wavelength (pixel index) domain. This does not create any problems because relative sensitivities among the channels are captured by the instrument functions as relative calibration factors, while $\alpha$ in \refeq{eq:for_alpha} takes care of the absolute calibration factor.

\begin{figure}
\includegraphics[width=\linewidth]{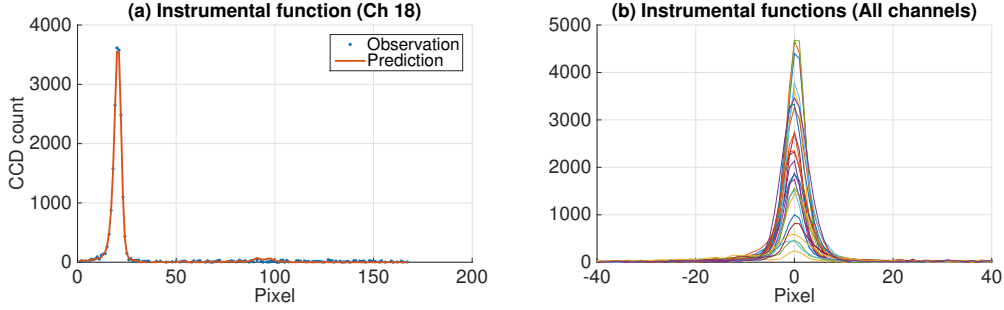}
\caption{(a) The observation (dots) and the MAP estimate of the instrument function (red line) for the channel 18. (b) The instrument functions (MAP) for all channels shown in different colours. Note that the instrument functions are not normalised in order to capture the relative sensitivities.}
\label{fig:instrument_effect}
\end{figure}

To infer the $i^\mathrm{th}$ channel instrument function $\mathbf{C}^i$, we use beam-into-gas shots. During the beam-into-gas shots, neutral lithium beam atoms are injected into the tokamak filled with a neutral deuterium gas whose pressure is less than $10^{-4}$ mbar. Because there is no plasma, there exists a negligible background signal caused by Bremsstrahlung ($B^i=0$). For this case, the posterior is $p\lp\mathbf{C}^i|\mathbf{D}^i\rp$ with
\begin{equation}
S^i_j=F^i_j\lp C^i_j A^i+B^i\rp+Z^i=F^i_j C^i_j A^i+Z^i, 
\label{eq:spectral-model-vector-instrument}
\end{equation}
where the interference filter function $\mathbf{F}^i$ is set to be the MAP estimation of $p\lp\mathbf{F}^i|\mathbf{D}^i\rp$ in \refeq{eq:bayes_spec}. Due to the small deuterium pressure inside the tokamak during the beam-into-gas experiments, a strong beam attenuation is not expected. According to \cite{10RSI_Brix}, there is no indication of any beam attenuation, so the emitted photons $N^{em}_{ph}$ should not vary along the beam. The variation of the observed intensities must therefore be due to differences in $T$, $Q$, and $\Delta z$ in \refeq{eq:for_alpha}. Assuming the Li I line emission is constant over the beam, $\mathbf{C}^i$ will give us these relative calibration factors. Since the electronic offset is not negligible for some channels as shown in \reffig{fig:instrument_effect}, we calculate posterior of both instrument function and offset $p\lp\mathbf{C}^i,Z^i|\mathbf{D}^i\rp$.

The likelihood $p\lp\mathbf{D}^i|\mathbf{C}^i,Z^i\rp$ is taken as the Gaussian with the mean given by \refeq{eq:spectral-model-vector-instrument}. We let the prior $p\lp\mathbf{C}^i\rp$ to have the form of \refeq{eq:prior-effect} with the covariance function \refeq{eq:cov-effect}. Again, the hyperparameters are set such that the evidence is maximised. The prior $p\lp Z^i\rp$ is a normal distribution with a zero mean and a very large variance ($10^6$). \reffig{fig:instrument_effect}(a) compares the observation and instrument function (MAP) for channel 18. \reffig{fig:instrument_effect}(b) shows the instrument functions (MAP) for all channels, which also capture the relative calibration factors.

\subsection{Line intensities}
\label{sec:inference-line}
We inferred $\mathbf{F}^i$ and $\mathbf{C}^i$ from \refsec{sec:inference-effect} and are left with three free parameters $A^i$, $B^i$ and $Z^i$ in \refeq{eq:spectral-model-vector}. The posterior $p\lp A^i,B^i,Z^i | \mathbf{D}^i \rp$ is calculated using a Gaussian likelihood $p\lp \mathbf{D}^i | A^i,B^i,Z^i \rp$ with the mean of $S^i_j=F^i_j\lp C^i_j A^i+B^i\rp+Z^i$. As we have three independent free parameters, the prior $p\lp A^i,B^i,Z^i \rp$ is
\begin{equation}
p\lp A^i, B^i, Z^i \rp = p\lp A^i\rp p\lp B^i\rp p\lp Z^i\rp,
\end{equation}
where all three priors are Gaussian distributions with a zero mean and very large variance ($10^6$).

\begin{figure}
\includegraphics[width=\linewidth]{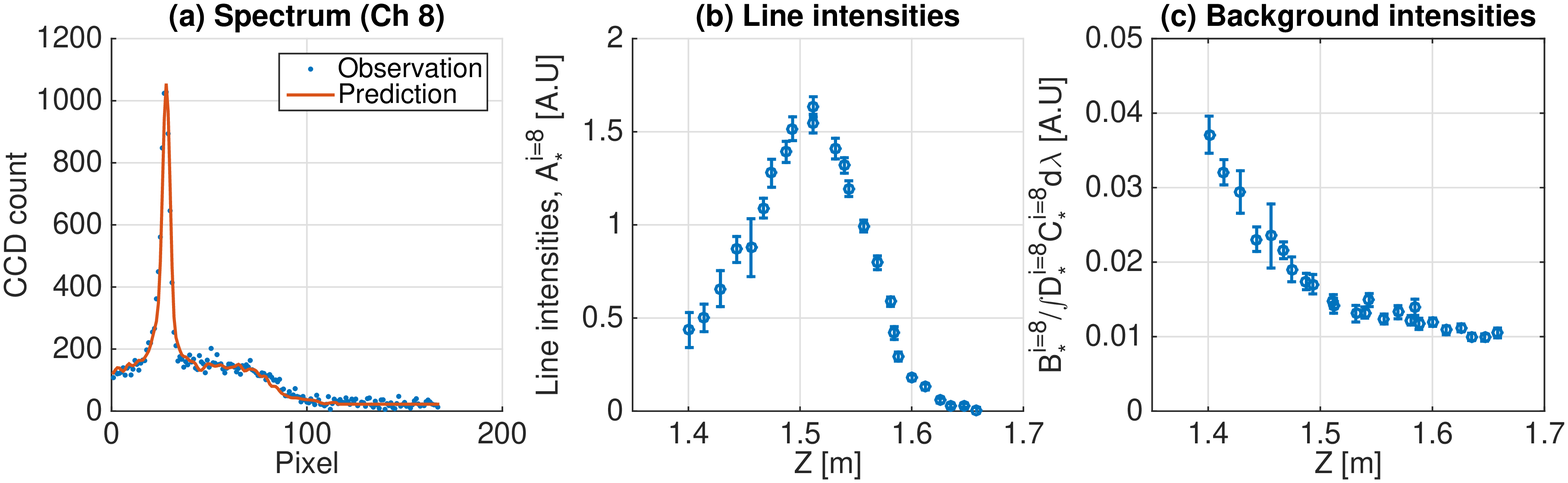}
\caption{(a) The measured spectrum (dots) and its MAP estimate at $50.260$ sec of the shot number $87861$ for channel 8. The profile of (b) Li I line and (c) background intensities (MAP) with their variances.}
\label{fig:spectral}
\end{figure}

A comparison between the Li I line and background intensities (MAP) and observation at $50.260$ sec of the JET shot number $87861$ for channel $8$ is shown in \reffig{fig:spectral}(a). \reffig{fig:spectral}(b) and (c) show the profiles of the Li I line and background intensities with their uncertainties (the shortest $95$ \% confidence interval), respectively. The edge $n_e$ profiles are directly inferred from the Li I line intensities. The profile of background radiation in \reffig{fig:spectral}(c) is most likely dominated by Bremsstrahlung emission, so could be used for inferring the effective charge $Z_{\mathrm{eff}}$, since Bremsstrahlung intensities are proportional to $Z_{\mathrm{eff}}$ ($I_\mathrm{Brem}\propto Z_{\mathrm{eff}} n_e^2 T_e^{1/2}$).

\subsection{Edge electron density profiles}
\label{sec:inference-density}
To infer the electron density profile, we take the MAP estimate of the Li I line intensities with their variances ($\mathbf{A}\pm\mathbf{\sigma_A}$). The posterior is given by 
\begin{equation}
p\lp \mathbf{n_{e}}, \alpha | \mathbf{A},  \mathbf{\sigma_A}\rp \propto p\lp\mathbf{A} | \mathbf{\sigma_A},\mathbf{n_{e}},\alpha \rp p\lp \mathbf{n_{e}}, \alpha\rp,
\label{eq:posterior-ne}
\end{equation}
where the absolute calibration factor $\alpha$ and the edge electron density profile $\mathbf{n_{e}}$ are the free parameters.

The likelihood $p\lp\mathbf{A} | \mathbf{\sigma_A},\mathbf{n_{e}}, \alpha \rp$ is given by
\begin{eqnarray}
p\lp\mathbf{A} | \mathbf{\sigma_A},\mathbf{n_{e}}, \alpha \rp = \frac1{\sqrt{\lp2\pi\rp^{N_{\mathrm{ch}}}\abs{\check\mathbf\Sigma_{A}}}}\exp{\lsb-\frac12 \lp  \mathbf{A} - \alpha \mathbf{N}_{2}  \rp^T \check\mathbf\Sigma_{A}^{-1}\lp \mathbf{A} - \alpha \mathbf{N}_{2}  \rp\rsb},
\nonumber\\
\label{eq:likelihood-ne}
\end{eqnarray}
where $N_{\mathrm{ch}} = 26$ is the total number of the channels. $\check\mathbf\Sigma_{A}$ is the $N_{\mathrm{ch}} \times N_{\mathrm{ch}}$ diagonal matrix with the entry of $\lp\sigma_A^i\rp^2$ at the $i^\mathrm{th}$ row and $i^\mathrm{th}$ column. We calculate $ \mathbf{N}_{2}$ using the Runge-Kutta method (RK4) from the model \refeq{eq:multi-state} with the initial condition \refeq{eq:initial}. 

We give $n_e$ and $\alpha$ independent priors, where $p\lp \alpha \rp$ is uniform between $1$ and $1000$. For $p\lp \mathbf{n_{e}}\rp$, based on a large database of existing profiles, we can estimate the hyperparameters for the Gaussian process prior. From this we set the hyperparameters $\sigma_f$ and $\ell$ for the covariance matrix $\check\mathbf{K}$ to be $20.0$ and $0.025$, respectively. We note that these values for the hyperparameters are not rigorously obtained by maximising the evidence due to the requirement of too much computation time. Nevertheless, these values give good fit to the data. A possible improvement would be to marginalise over these hyperparameters as in \cite{11EFDA_Svensson}.

\begin{figure}
\includegraphics[width=\linewidth]{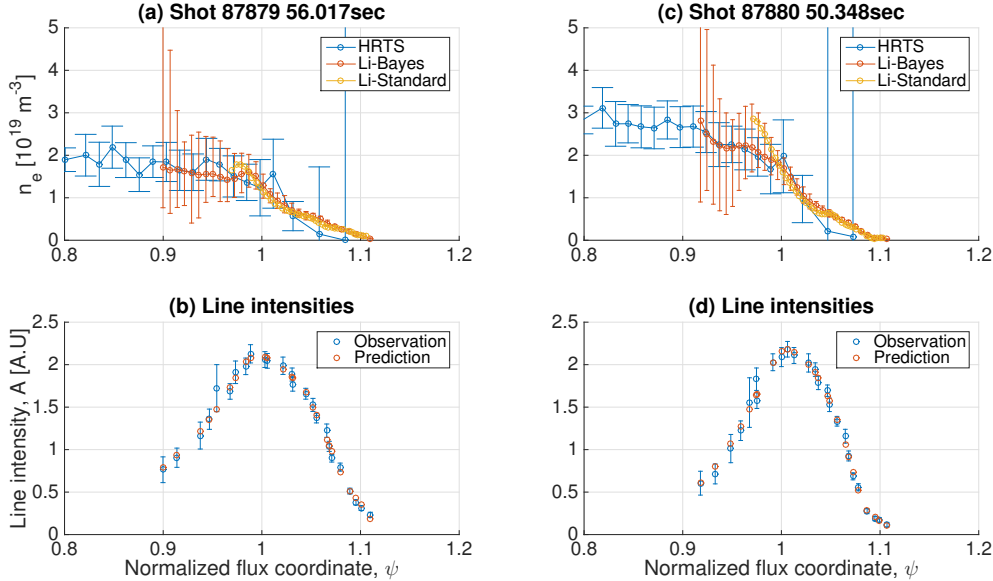}
\caption{ (a) The MAP estimate of the edge electron density profile (red) and the associated uncertainties (shortest 95\% interval) together with the $n_e$ profiles from the HRTS system (blue) and conventional Li-BES analysis (yellow). (b) the MAP estimate of the Li I intensities (blue), i.e., $\mathbf{A}$ in \refeq{eq:posterior-ne}, and the prediction (red), i.e., $\alpha \mathbf{N}_{2} \lp\mathbf{n_{e}}\rp$ for the shot \#87879 at $56.017$ sec. (b) and (d) are same as (a) and (c) for the shot \#87880 at $50.348$ sec.}
\label{fig:results}
\end{figure}

The posterior of $\mathbf{n_{e}}$ and $\alpha$ is explored by a Markov Chain Monte Carlo (MCMC) sampling scheme. \reffig{fig:results}(a) and (c) show the MAP estimate of the edge electron density profiles (red) with their associated uncertainties, which cover $95$\% of the samples from posterior, i.e., the shortest 95\% interval. For the sake of comparison, $n_e$ profiles from the HRTS system (blue) and results from the conventional analysis of the JET Li-BES system (yellow) \cite{10RSI_Brix, 12RSI_Brix} are also shown in the same figures. \reffig{fig:results}(b) and (d) show the MAP estimates of the Li I line intensities from the previous section (blue), i.e., $\mathbf{A}$ in \refeq{eq:posterior-ne}, and prediction (red), i.e., $\alpha \mathbf{N}_{2}$, for \reffig{fig:results}(a) and (c), respectively. 

It is clear from these results that we have inferred a proper absolute calibration factor $\alpha$ even though we have not used the singular point method \cite{92PPCF_Schweinzer}. The range of the density profile inference has been extended to the full observation range which was not possible with the conventional data analysis method. We stress that we have not used a separate background measurement via Li neutral beam modulations  because our method is capable of providing intensities of Li I line and background radiations simultaneously. Finally, we also have not made an assumption of monotonic profile, either.

\begin{figure}
\includegraphics[width=\linewidth]{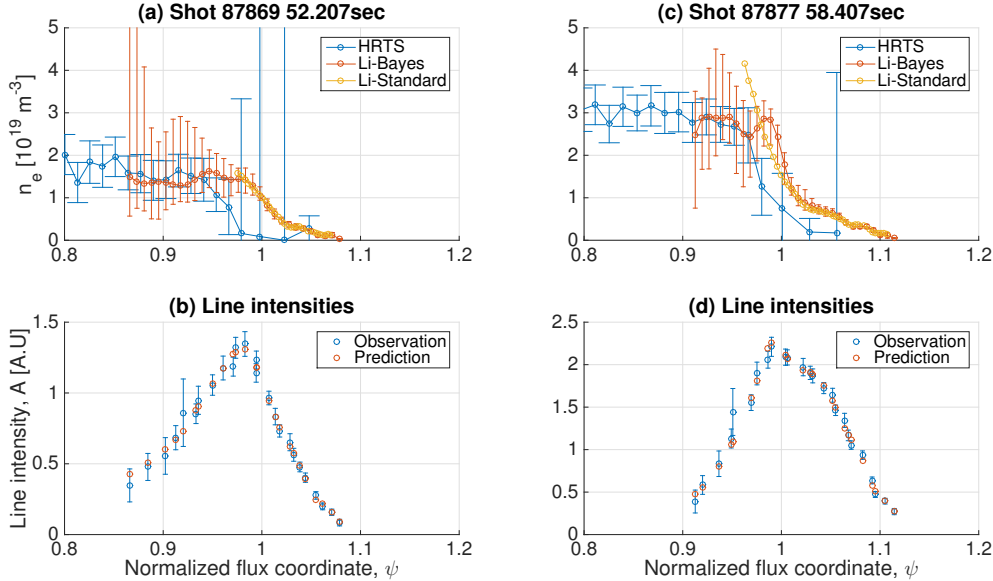}
\caption{Same as \reffig{fig:results} for a different time and shot number, showing disagreement between the Li-BES and HRTS analysis although the prediction of the Li line intensities matches well with their MAP estimate.}
\label{fig:shifted}
\end{figure}

In some cases, we observe a difference between the profiles inferred from the Li-BES and HRTS systems (\reffig{fig:shifted}). Calibration of the spatial position for the Li-BES may be questioned. However, this calibration is performed with relatively high reliability \cite{10RSI_Brix}. We do suspect that it may have been caused by the EFIT reconstruction. The Li-BES system injects neutral lithium beam atoms vertically from the top of the JET at major radius $R=3.25$ m and covering the vertical position $Z=1.67\sim1.40$ m approximately; whereas the HRTS system observes electron density along the laser penetrating horizontally at the midplane ($R=2.9\sim3.9$ m and $Z=0.06\sim0.11$ m). The flux coordinate mapping provided through EFIT may well be inaccurate when comparing the midplane with the top of the vessel. We leave further investigation of this issue to future work.

In \reffig{fig:results}(a) and (c) and \reffig{fig:shifted}(a) and (c) we can see that the uncertainties of the electron densities in the inner region is larger than those of the outer region. This result cannot be explained solely by the number of detected photons as attested by \reffig{fig:results}(b) and (d) and \reffig{fig:shifted}(b) and (d). This trend of larger uncertainties in the inner region is also observed in ASDEX Upgrade \cite{08PPCF_Fischer, 14PPCF_Willensdorfer}. Here, we provide two qualitative reasons to explain this trend. As shown in \reffig{fig:atomicdata}, the relative population of the first excited state $N_2$ becomes less sensitive to the change of $n_e$ as it increases. Typically, $n_e$ is larger in the inner region than the outer region, therefore the similar level of uncertainty in $N_2$ corresponds to a larger uncertainty of $n_e$ in the inner region. In addition, the neutral Li beam attenuation as it penetrates into the plasmas can cause this trend of increasing uncertainties: consider two separate measurements of the absolute number of the first excited state which both give the same value of $200\pm20$ where the total number of neutral beam atoms is $500$ in one case and $1000$ in another case. Then, the relative population $N_2$ is $(200\pm20)/500=0.4\pm0.04$ for the former case and $(200\pm20)/1000=0.2\pm0.02$ for the latter case. It is evident that the former case has the larger uncertainty than the latter case even if the absolute numbers of the first excited state are the same for both cases. Therefore, the beam attenuation, i.e., decrease of the total number of beam atoms, can cause the larger uncertainty of $n_e$ in the inner region \cite{14PPCF_Willensdorfer}. Finally, we note that there can be additional effects from the uncertainties of the absolute calibration factor \cite{92PPCF_Schweinzer, 93PPCF_Pietrzyk}.

\section{Conclusion}
\label{sec:conclusion}
In this paper, we have presented a Bayesian model to obtain edge electron density profiles based on the measured JET Li-BES spectra. The model has been implemented in the Minerva Bayesian modelling framework. Our scheme includes uncertainties due to photon statistics and electric noise estimated from the measured data obtained with the transmission grating spectrometer. The instrument effects such as the interference filter function and instrument function are inferred from separate measurements using Gaussian processes whose hyperparameters are selected by evidence maximisation. Also the electron density profiles are modelled using Gaussian processes, whose hyperparameters are determined from the JET historical electron density  profiles. Inference is done through maximisation of the posterior (MAP) and Markov Chain Monte Carlo Method (MCMC) sampling. The Li I line and background intensities are simultaneously inferred as well as their associated uncertainties, thereby eliminating extra effort of measuring background intensity via Li neutral beam modulations.

\section{Acknowledgement}
This work is supported by National R\&D Program through the National Research Foundation of Korea (NRF) funded by the Ministry of Science, ICT and Future Planning (Grant No. 2014M1A7A1A01029835) and the KUSTAR-KAIST Institute, KAIST, Korea. This work has been carried out within the framework of the EUROfusion Consortium and has received funding from the Euratom research and training programme 2014-2018 under Grant Agreement No. 633053. The views and opinions expressed herein do not necessarily reflect those of the European Commission.

\section*{References}
\bibliography{NF2015}

\end{document}